\documentclass[prb,twocolumn,showpacs,superscriptaddress,floatfix]{revtex4-1}
\usepackage{graphicx,amsfonts,amssymb,amsmath}
\usepackage[pagebackref=true,colorlinks,linkcolor=blue,citecolor=red]{hyperref}

\newcommand{\be}{\begin{equation}}
\newcommand{\ee}{\end{equation}}
\newcommand{\bearr}{\begin{eqnarray}}
\newcommand{\eearr}{\end{eqnarray}}
\newcommand{\rr}{r(q)}
\newcommand{\rre}{R(q)}

\begin{document}

\title{One directional Polarized Neutron Reflectometry with optimized reference layer method}

\author{Seyed Farhad Masoudi}
\email{masoudi@kntu.ac.ir}
%\affiliation{Department of Physics, K.N. Toosi University of Technology, P.O. Box 15875-4416, Tehran, Iran}

\author{Saeed S. Jahromi}
\email{s.jahromi@dena.kntu.ac.ir}
\affiliation{Department of Physics, K.N. Toosi University of Technology, P.O. Box 15875-4416, Tehran, Iran}

\begin{abstract}

  In the past decade, several neutron reflectometry methods for determining the modules and phase of complex reflection coefficient of an
unknown multilayer thin film  have been worked out among which the method of variation of surroundings and reference layers  are of highest 
interest. These methods were later modified for measurement of polarization of 
reflected beam instead of the measurement of intensities. In their new architecture, these methods not only suffered from the necessity 
of change of experimental setup, but also another difficulty was added
to their experimental implementations which was related to the limitations of the technology of the neutron reflectometers that could only measure 
the polarization of the reflected neutrons in the same direction as the polarization of the incident beam. As the instruments are limited, the 
theory has to be optimized so that the experiment could be performed. In a recent work, we have developed the method of variation of surroundings 
for one directional polarization analysis. In this new work, the method of reference layer with polarization analysis has been optimized to 
determine the phase and modules of the unknown film with measurement of the polarization of reflected neutrons in the same direction as the polarization
of the incident beam.
\end{abstract}
\pacs{61.12.Ha Neutron reflectometry}
\maketitle

\section{Introduction}

As an atomic scale probe, neutron reflectometry with polarized neutrons has a vast application to the study of layered media and thin films. 
Each material has a unique scattering length density (SLD) for neutrons which can be used as a unique characteristic for determining 
the type and thickness of that material \cite{Xiao-Lin}. 
In other words, retrieving the nuclear scattering potential of neutrons from reflectivity data can be used as a powerful tool for identifying the 
unknown samples. However, measuring the intensity of the reflected neutrons from the sample would not lead to a unique result for the SLD of the sample.
In the scattering process, the reflectivity $\rre =\mid\rr\mid^2$ is measured and the information of the phase of the complex reflection 
coefficient $\rr$  is lost. 
As a result, two different layer with different phase information can have the same reflectivity.
Hence, the full knowledge of the phase of complex reflection coefficient is necessary in order to retrieve a unique result from the 
reflectometry experiments \cite{Xiao-Lin, Majkrzak1}. 

In the paste decades, several theoretical methods for resolving this so-called \textit {phase problem} have been worked out such as
variation of surroundings medium \cite{Majkrzak2,Majkrzak3} which makes use of the
controlled variations of scattering length density of the incident and/or substrate medium or the 
method of reference layers \cite{Majkrzak4,Majkrzak5} which is based on the interference between 
the reflections of a known reference layer and the unknown surface profile.
Either of the methods suffer from the experimental difficulties regarding the change of 
the substrate or reference layers for each reflectivity measurement \cite{Majkrzak3, Majkrzak6}.
This inefficiency was recovered by using polarized neutrons and magnetic layers as substrate \cite{Masoudi1,Masoudi2}
or reference layer \cite{Leeb1,Masoudi3,Masoudi4}.
Later on, these methods were enhanced to measure the polarization 
of the reflected beam to determine the phase of reflection\cite{Leeb2}.
Although the polarization based approaches had been theoretically proven to be efficient, their experimental 
implementation was limited by the ability of reflectometers in the polarization measurement direction. 
As these methods required at least two measurements
of polarization of the reflected bean in different directions \cite{Leeb2,Masoudi3,Masoudi4}, their experimental implementations
were not practical with reflectometers which can only measure the polarization
of the reflected neutrons in the same direction as the incident beam.

In a recent work \cite{jahromi}, we developed the method of variation of surroundings with one directional polarization
analysis and discussed about possibility of experimental implementation of the method. 
In this new worked, we have optimized the method of reference layers
based on the measurement of the polarization of reflected beam in
the same direction as the incident neutrons.

In section~\ref{method}, the foundation of the optimized reference layer is formulated. Section \ref{example}, deals
with numerical examination of the method for an unknown sample. 
The SLD of the sample is also retrieved from the phase information which was obtained from optimized
reference layer method. The paper ends with further discussions about the experimental challenges of the method. 

\section{Optimized reference layer method}
\label{method}

Scattering of neutrons from a sample is described by optical potential, $v(z)=2\pi\hbar^2\rho(x)/m$, 
where $\rho(x)=\rho_n\pm\rho_m$ is the scattering length density
of the sample as a function of its depth. $\rho_n$ is the nuclear part of the SLD and
$\rho_m=(m/2\pi\hbar^2)\mu B$ is the magnetic part. The Plus (minus) signs denotes
the incident beam polarized parallel (anti parallel) to the local magnetization. 
The scattering of neutrons from the sample is described by the one dimensional 
Schrödinger equation: 
\be \label{eq1}
[\partial_{x}^2+(q^2-4\pi\rho(x))]\psi(q,x)=0
\ee
where $q$ is the incident neutron wave number in the $x$ direction.
The reflection ($r_{\pm}$) and transmission ($t_{\pm}$) coefficients of the sample can be
determined from Eq.~\ref{eq1} using the transfer matrix method \cite{Majkrzak1}:
\be \label{eq2}
\begin{pmatrix}
1 \\
ih \\
\end{pmatrix} t_{\pm}e^{ihqL}=
\begin{pmatrix}
A(q) & B(q) \\
C(q) & D(q) \\
\end{pmatrix}
\begin{pmatrix}
1+r_{\pm} \\
if(1-r_{\pm}) \\
\end{pmatrix}
\ee
where $L$ is the thickness of the sample and $j=(1-4\pi\rho_{j}/q^2)^{1/2}$ with $j=f,h$ is the refractive 
index of the fronting and backing medium respectively 
($\rho_f$ and $\rho_h$ are the SLDs of the fronting and backing medium).
$(A,\ldots, D)$ are the elements of the transfer matrix which are uniquely determined
as a function of the SLD of the sample. The reflection coefficient of 
the sample in terms of the transfer matrix elements is also written as follows:
\be \label{eq3}
r_{\pm}^{fh}=\frac{\beta_{\pm}^{fh}-\alpha_{\pm}^{fh}-2i\gamma_{\pm}^{fh}}{\beta_{\pm}^{fh}+\alpha_{\pm}^{fh}+2}
\ee
where 
\be \label{eq4}
\begin{gathered}
\alpha_{\pm}^{fh}=f^{-1}hA^2+f^{-1}h^{-1}C^2 \\
\beta_{\pm}^{fh}=fhB^2+fh^{-1}D^2 \\
\gamma_{\pm}^{fh}=hAB+h^{-1}CD 
\end{gathered}
\ee
The reflectivity depends on $\alpha_{\pm}$, $\beta_{\pm}$ and $\gamma_{\pm}$ in terms of a new quantity $\Sigma_{\pm}$: 
\be \label{eq5}
\Sigma_{\pm}^{fh}(q)=2 \frac{1+R_{\pm}^{fh}}{1-R_{\pm}^{fh}}=\beta_{\pm}^{fh}+\alpha_{\pm}^{fh} \\
\ee
\begin{figure}
\centering{\includegraphics[width=\columnwidth]{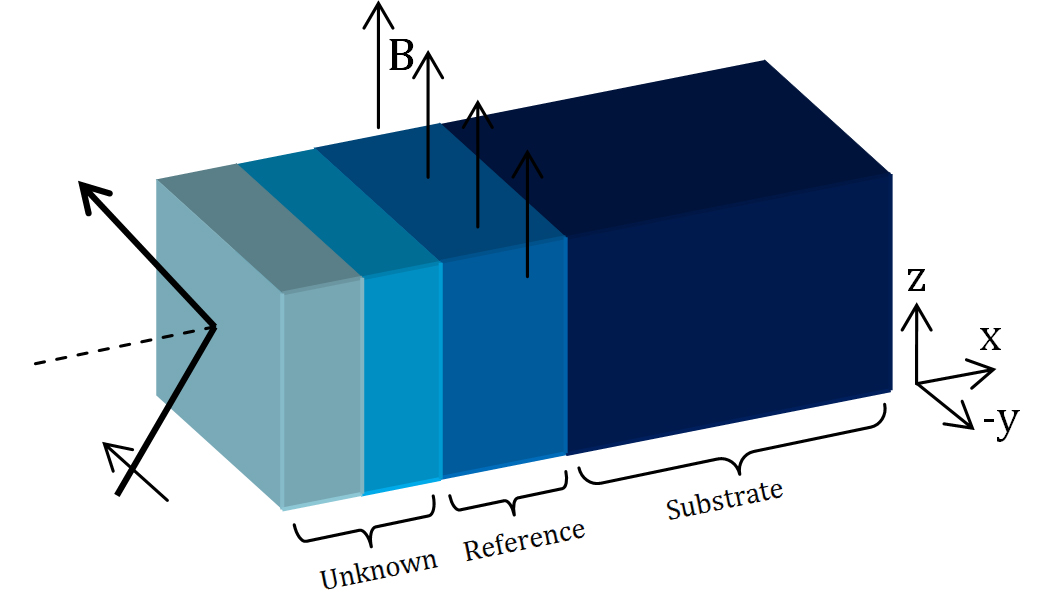}} \caption{Pictorial demonstration of 
the arrangement of the sample. An unknown bilayer film is mounted on top
of the known magnetic reference layer which is deposited over an infinite substrate} 
\label{fig1}
\end{figure}
Suppose the sample is composed of two part, an unknown part which is mounted on top of the known magnetic reference layer
which is magnetized in the $+z$-direction
(Fig.~\ref{fig1}).
The transfer matrix of the whole sample is written as the multiplication of the transfer matrix of each part:
\be \label{eq6}
\begin{pmatrix}
A & B \\
C & D \\
\end{pmatrix}=
\begin{pmatrix}
w_{\pm} & x_{\pm} \\
y_{\pm} & z_{\pm} \\
\end{pmatrix}
\begin{pmatrix}
a & b \\
c & d \\
\end{pmatrix}
\ee
where $(a \dots d)$ is for the unknown part of the sample and $(w \dots z)$ belong to the known part.
Using Eq.~\ref{eq5},\ref{eq6} , $\Sigma_{\pm}$ is modified as follows: 
\be \label{eq7}
\Sigma_{\pm}^{fh}=\beta_{k\pm}^{fh}\widetilde{\alpha}_{u}^{ff} + \alpha_{k\pm}^{fh} \widetilde{\beta}_{u}^{ff} + 2\gamma_{k\pm}^{fh}\widetilde{\gamma}_{u}^{ff}
\ee
where the subscript $k$ ($u$) denotes the known (unknown) part of the sample and the tilde
represents the mirror-reversed sample which is obtained by the interchange
of the diagonal elements of the transfer matrix, ($A\leftrightarrow D$). 
The superscript $ff$ denotes a sample with the same medium at both sides with refractive index $f$.

The polarization of the reflected beam, ($p_x, p_y, p_z$), is written in terms of the polarization of the incident 
beam, ($p_{x}^{0}, p_{y}^{0}, p_{z}^{0}$) \cite{Leeb1, Gottfried}:
\be \label{eq8}
p_x + i p_y= \frac{2(r_{+}^{fh})^{*} r_{-}^{fh} (p_{x}^{0} +i p_{y}^{0})}{R_{+}^{fh} (1+p_{z}^{0}) + R_{-}^{fh} (1-p_{z}^{0})}
\ee
Up to now, we have defined all the preliminary parameters. Here, we are going to perform two
different polarization measurements which satisfy our goals:

{\bf First measurement:} We suppose the incident neutrons to be fully polarized in the $y$ direction. 
The polarization of the reflected beam, $P_{1}$, is also 
measured in the same direction as the incident beam. From Eq.~\ref{eq8} for the measured polarization, we have:
\be \label{eq9}
P_{1}=1-\frac{2\xi_{k}}{\Sigma_{+}^{fh} \Sigma_{-}^{fh}-4}
\ee
where
\be \label{eq10}
\xi_{k}=\alpha_{k+}^{fh}\beta_{k-}^{fh} + \beta_{k+}^{fh}\alpha_{k-}^{fh} - 2 (1+\gamma_{k+}^{fh}\gamma_{k-}^{fh} )
\ee
which is completely determined from the knowledge of the known part of the sample.

{\bf Second measurement:} For the second case, we consider a non-polarized incident beam, 
rotate the sample by $90^{\circ}$ in the $y-z$ plane and measure
the polarization of the reflected neutrons in the the $y$ direction. 
This measurement is the same as measuring the polarization in the direction of the local
magnetization of the reference layer, $z$, for the non-rotated sample. 
This choice for the polarization analysis direction would not contradict with our first 
assumption of fixed experimental setup. From Eq.~\ref{eq8}, the second polarization can be written as:
\be \label{eq11}
P_{2}=\frac{2(\Sigma_{+}^{fh}-\Sigma_{-}^{fh})}{\Sigma_{+}^{fh} \Sigma_{-}^{fh}-4}
\ee

As the knowledge of $P_{1}$ and $P_{2}$ are experimentally known, $\Sigma_{+}^{fh}$ is determined from the following quadratic equation:
\be \label{eq12}
(\Sigma_{+}^{fh})^2-\frac{\xi_{k}P_{2}}{1-P_{1}} \Sigma_{+}^{fh} - (4 + \frac{2\xi_{k}}{1-P_{1}})=0
\ee
This quadratic equation has two different solutions; the physical solution is selected
using the fact that $\Sigma_{+} > 2$. By knowing $\Sigma_{+}^{fh}$, the $\Sigma_{-}^{fh}$ is also determined as follows:
\be \label{eq13}
\Sigma_{-}^{fh}=\Sigma_{+}^{fh} - \frac{\xi_{k}P_{2}}{1-P_{1}} 
\ee

Using Eq.~\ref{eq7} and the knowledge of $\Sigma_{\pm}^{fh}$, the parameters of the unknown part of the sample are related to each other as follows:
\be \label{eq14}
\begin{gathered}
\widetilde{\beta}_{u}^{ff}=a_{1} \widetilde{\gamma}_{u}^{ff} + b_{1} \\
\widetilde{\alpha}_{u}^{ff}=a_{2} \widetilde{\gamma}_{u}^{ff} + b_{2}
\end{gathered}
\ee
where
\be \label{eq15}
\begin{gathered}
a_{1}=-2 \frac{\gamma_{k+}^{fh}\beta_{k-}^{fh} - \beta_{k+}^{fh}\gamma_{k-}^{fh}}{\alpha_{k+}^{fh}\beta_{k-}^{fh} - \beta_{k+}^{fh}\alpha_{k-}^{fh}} \\
a_{2}=-2 \frac{\gamma_{k+}^{fh}\alpha_{k-}^{fh} - \alpha_{k+}^{fh}\gamma_{k-}^{fh}}{\alpha_{k-}^{fh}\beta_{k+}^{fh} - \beta_{k-}^{fh}\alpha_{k+}^{fh}} \\
b_{1}=\frac{\Sigma_{k+}^{fh}\beta_{k-}^{fh} - \Sigma_{k-}^{fh}\beta_{k+}^{fh}}{\alpha_{k+}^{fh}\beta_{k-}^{fh} - \beta_{k+}^{fh}\alpha_{k-}^{fh}} \\
b_{2}=\frac{\Sigma_{k+}^{fh}\alpha_{k-}^{fh} - \Sigma_{k-}^{fh}\alpha_{k+}^{fh}}{\alpha_{k-}^{fh}\beta_{k+}^{fh} - \beta_{k-}^{fh}\alpha_{k+}^{fh}} \\
\end{gathered}
\ee

By using Eq.~\ref{eq14} and the fact that $\gamma=\sqrt{\alpha \beta -1}$ \cite{Majkrzak5}, the $\widetilde{\gamma}_{u}^{ff}$ parameter is determined from the following quadratic equation:
\be \label{eq16}
(a_{1}a_{2}-1)\widetilde{\gamma}^{2}+(a_{1}b_{2}+a_{2}b_{1}) \widetilde{\gamma} +(b_{1}b_{2}-1)=0
\ee
The complex reflection coefficient is then determined as a function of $\widetilde{\gamma}_{u}^{ff}$:
\be \label{eq17}
\widetilde{r}_{u}^{ff}=\frac{(a_{1}-a_{2})\widetilde{\gamma}_{u}^{ff} + (b_{1}-b_{2})-2 i \widetilde{\gamma}_{u}^{ff}}{(a_{1}+a_{2})\widetilde{\gamma}_{u}^{ff} + (b_{1}+b_{2})+2}
\ee
The quadratic equation \ref{eq16} has two different solutions which only one of them is physical.
The physical solution can not directly be determined from Eq.~\ref{eq16}. In order to determine the physical solution, we put the two 
solutions into Eq.~\ref{eq17}
and select the physical one from the behavior of the reflection coefficient:
\be \label{condition1}
0<Re(r)<1, \	\  0<Im(r)<1
\ee
\be\label{condition2}
Re(r)|_{q\rightarrow 0} \longrightarrow -1, \	\ Im(r)|_{q\rightarrow 0} \longrightarrow 0  
\ee
As both of the two solutions satisfy Eq.~\ref{condition1}, the physical solution should be determined 
from their behavior in $q\rightarrow 0$ (Eq.~\ref{condition2}). 
When the physical solution in $q\rightarrow 0$ is determined, the solution in larger $q$ values 
is also determined based on the continuity of the reflection coefficient along the whole range of $q$ values.

\section{Numerical Example}\label{example}

In this section, we numerically examine the optimized reference layer method to certify the reliability of the method. 
We consider $(NiO$/$Fe_{2}O_{3})$
bilayer as the unknown sample which is mounted on top of a Cobalt layer as magnetic reference.
The whole sample is deposited over a silicon substrate. The SLD and thickness of
the layers are listed in table.~\ref{tab1}.
\begin{table}[!t]
 \begin{center}
 \caption{Information of the sample}
 \label{tab1}
   \begin{ruledtabular}
   \begin{tabular}{lccccc}
    Layer            &Thickness (nm)&$\rho_n(10^{-4}nm^{-2})$&$\rho_m(10^{-4}nm^{-2})$  \\
    \hline
    NiO              &20            &8.84		    &0  \\
    Fe$_{2}$O$_{3}$  &15	    &7.26		    &0  \\
    Co               &20   	    &2.23		    &4.21 \\
    Si               &Semi Infinite &2.08		    &0 \\
    \end{tabular}
  \end{ruledtabular}
 \end{center}
\end{table}
According to the the optimized reference layer method, for the first step, a $y$-polarized neutron
beam, $P=(0,1,0)$, was radiated to the sample and the polarization of
the reflected beam, $P_1$, was measured in the same direction. 
For the second measurement, first the sample is rotated by 90$^o$ in the $y-z$ plane and 
then by using a non-polarized incident beam, $P=(0,0,0)$, the polarization of the 
reflected beam, $P_2$, is measured in the $z$ direction.  
Using the measured polarizations in Eqs.~\ref{eq9}, \ref{eq11} and following 
the Eqs.~\ref{eq12}-\ref{eq17}, the complex reflection coefficient of the mirror image
of the unknown part of the sample while it is surrounded by vacuum at both sides, is determined. 
\begin{figure}
\centering{\includegraphics[width=\columnwidth]{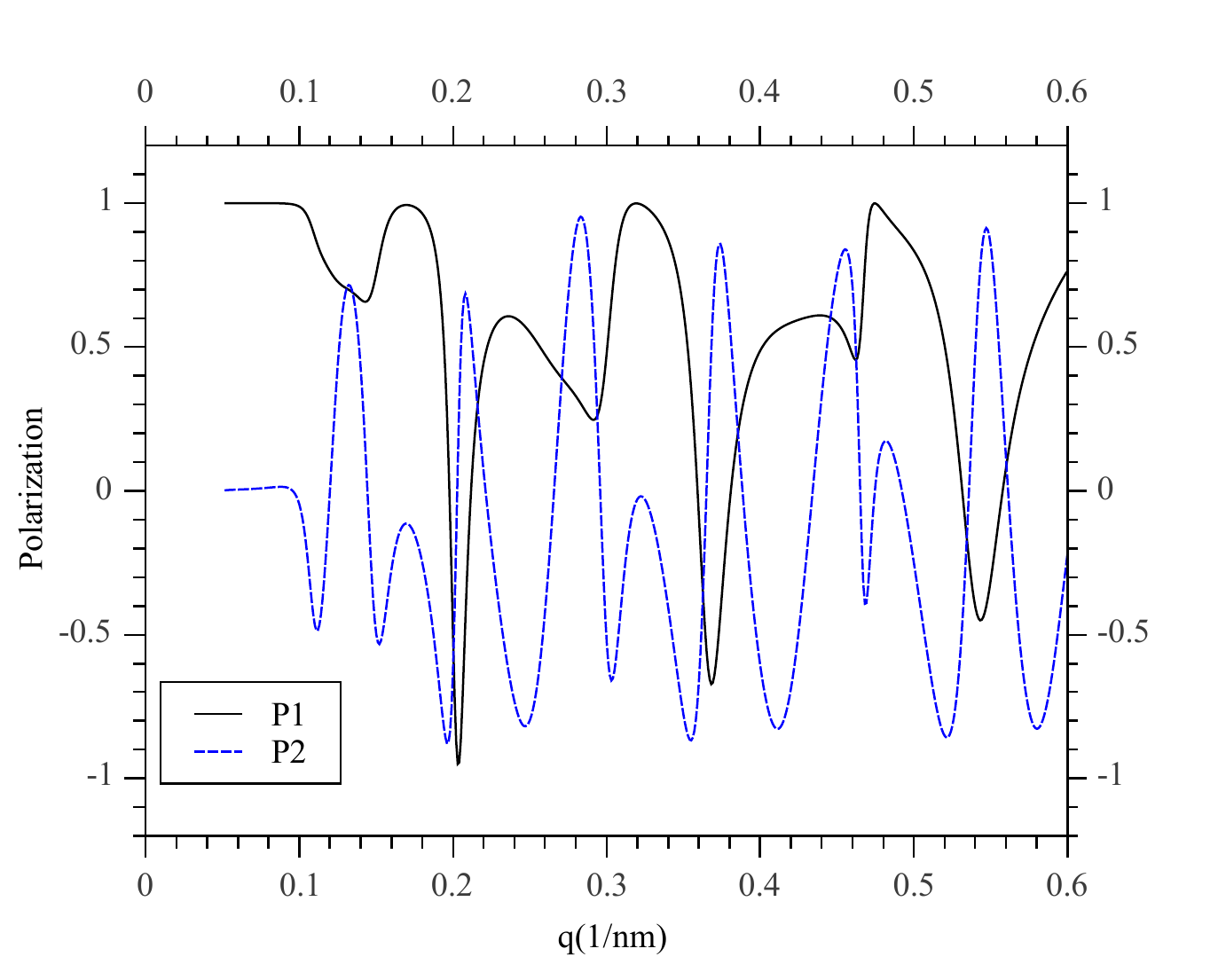}} 
\caption{Polarizations of the reflected neutrons for the two different measurements} 
\label{fig2}
\end{figure}

\begin{figure}
\centering{\includegraphics[width=\columnwidth]{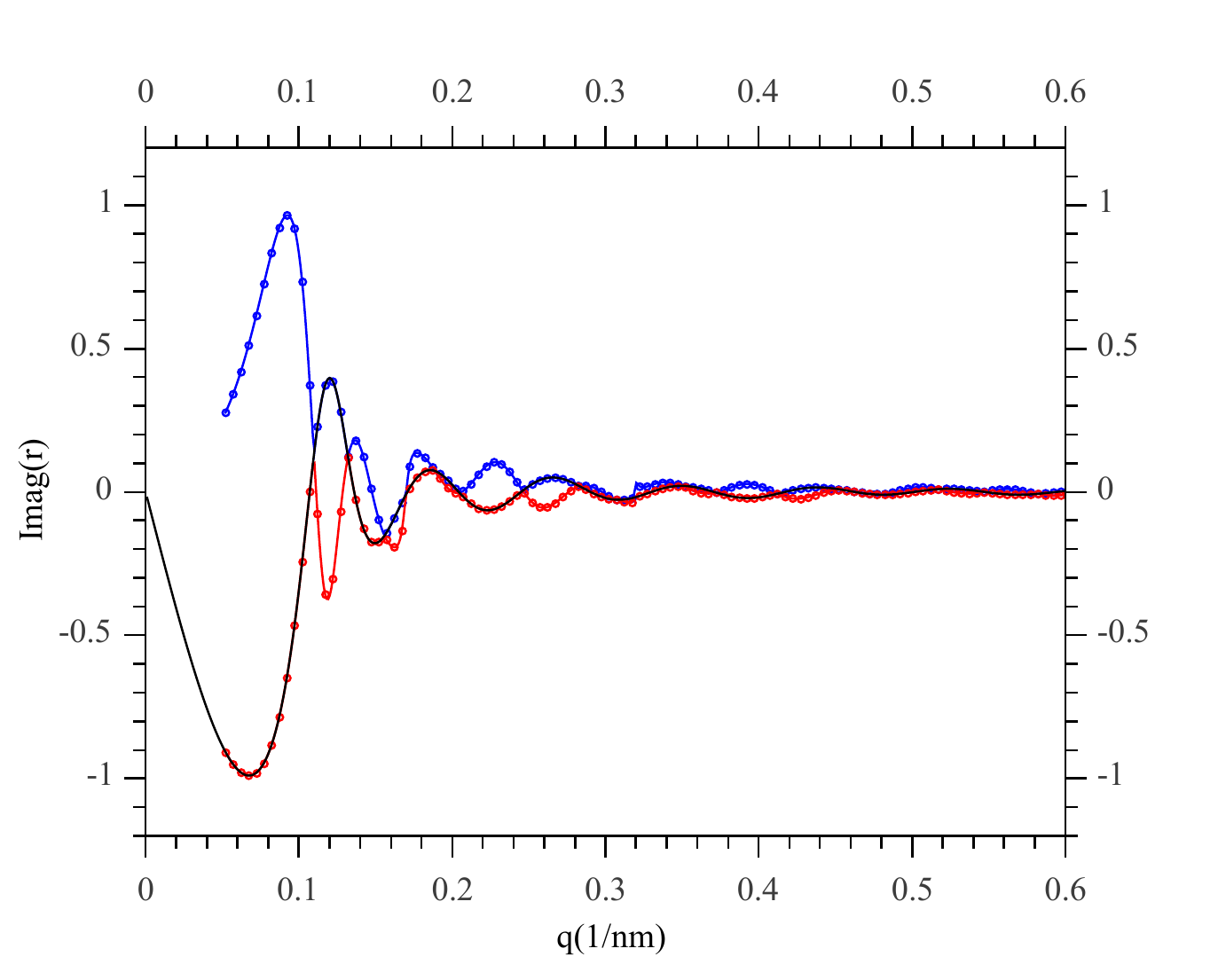}} 
\caption{Imaginary part of the complex reflection coefficient which is determined from Eq.\ref{eq17}. The physical solution alternates between the two different
solutions of the quadratic equation (blue and red curves). The solid black line shows the physical solution} 
\label{fig3}
\end{figure}

\begin{figure}
\centering{\includegraphics[width=\columnwidth]{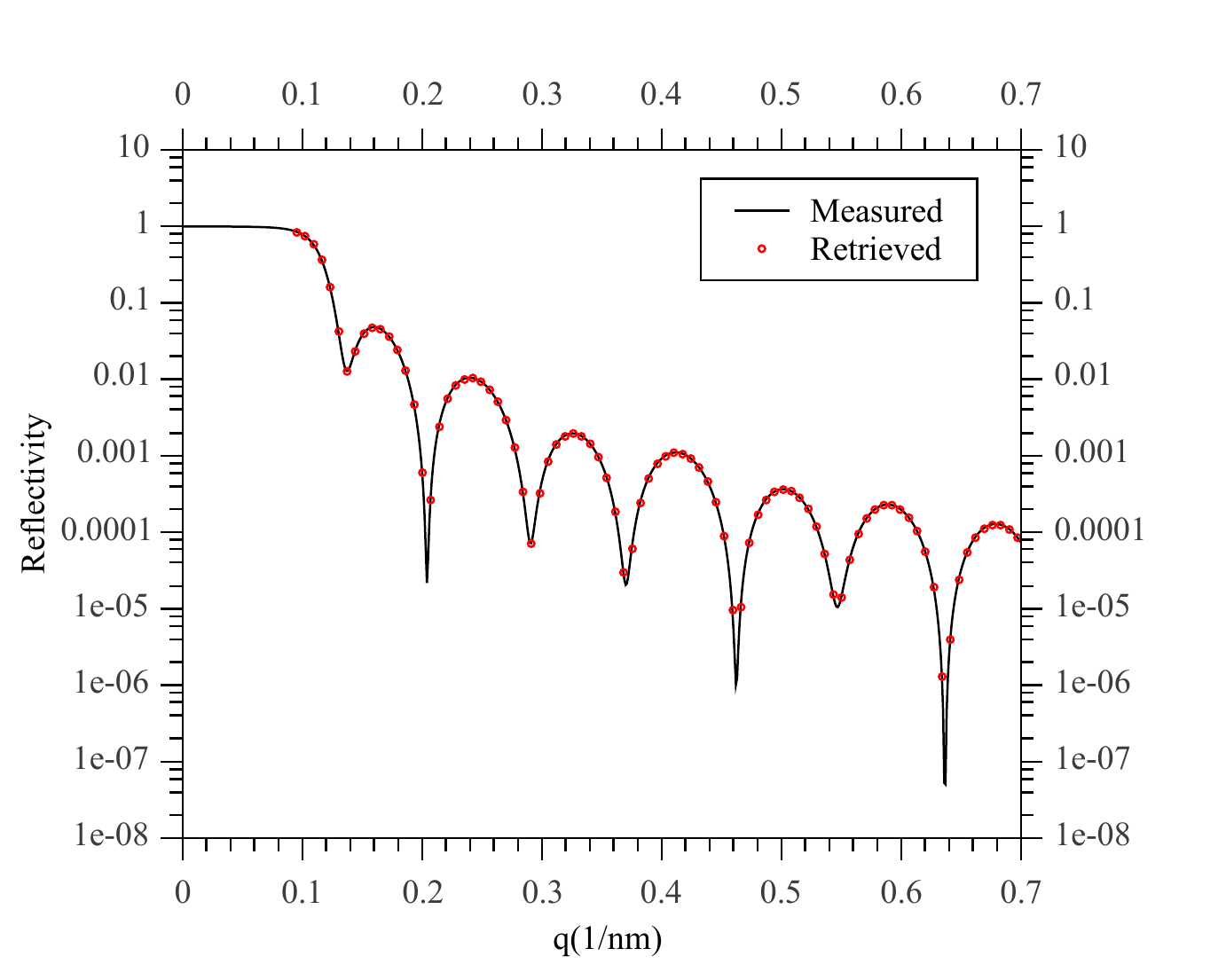}}
\caption{Reflectivity of the mirror reversed sample with vacuum surroundings. The retrieved reflectivity (red curve), clearly corresponds to the measured reflectivity} 
\label{fig4}
\end{figure}

Fig.~\ref{fig2} shows the polarization of the reflected beam for the two different measurements.
Fig.~\ref{fig3} also demonstrate the imaginary part of the complex
reflection coefficient of the sample which is retrieved from Eq.~\ref{eq17}. 
As it is illustrated in the figure, the physical solution alternates between these two
sets. The physical solution is selected based on the continuity of the data and the fact that as $q\longrightarrow 0$, 
the imaginary part of the complex reflection
coefficients, $Im(r)\longrightarrow 0$ from the negative \cite{Majkrzak3}. 
The retrieved reflectivity of the sample is also illustrated in Fig.~\ref{fig4} which
truly corresponds to the measured reflectivity from the experiment. 

\begin{figure}
\centering{\includegraphics[width=\columnwidth]{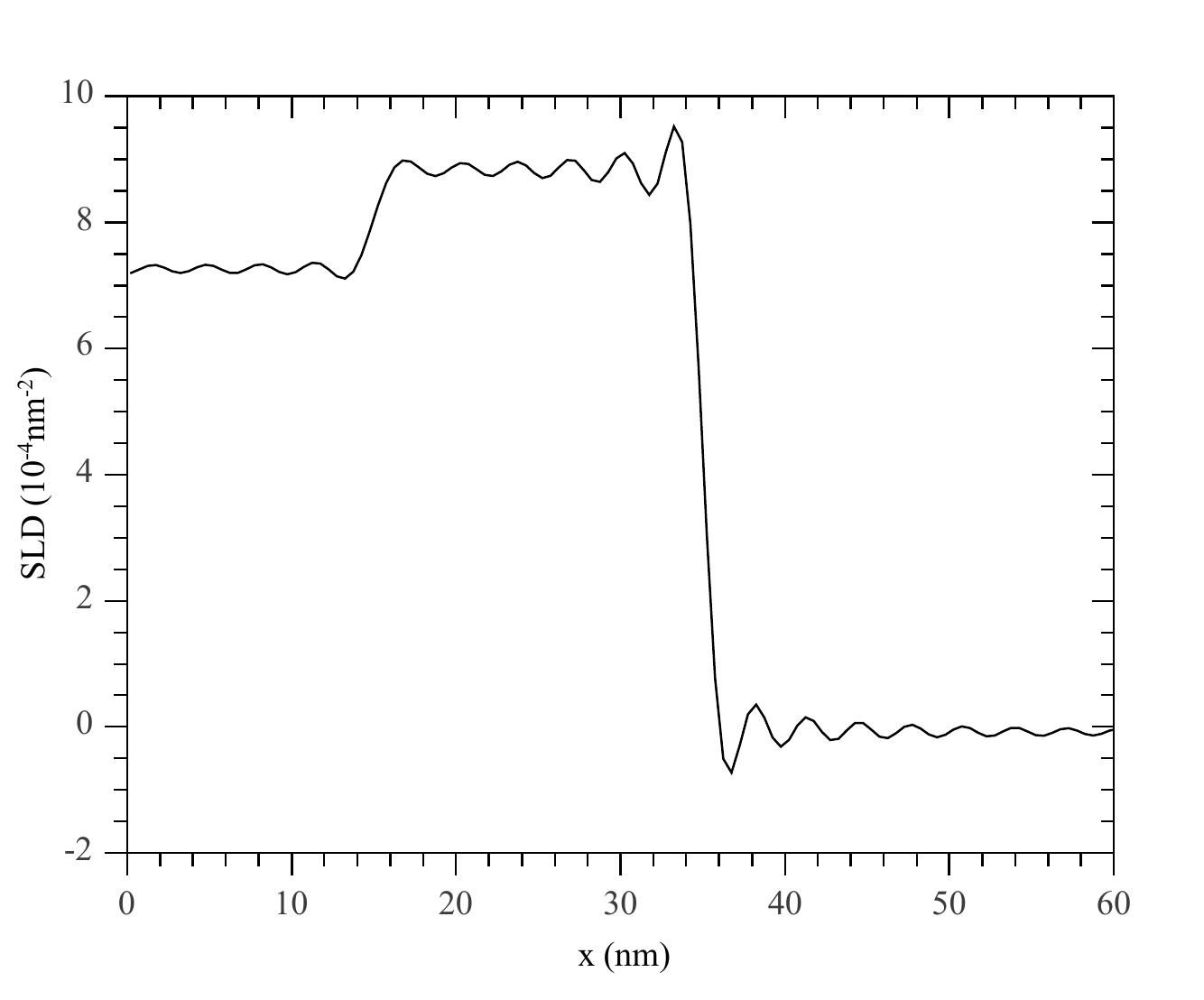}}
\caption{Retrieved Optical Potential $v$ of the sample from the data of Eq.\ref{eq17} and using the Sacks Code \cite{Sacks}.} 
\label{fig5}
\end{figure}

Knowing the information of the complex phase of reflection, the SLD of the sample can be determined
by solving the inverse Schrödinger equation for the scattering
potential. In neutron reflectometry problems, the SLD of the sample is retrieved by solving the  
Gel’fand–Levitan–Marchenko integral equations \cite{Majkrzak1, Majkrzak2, Chadan}. Sacks et al. \cite{Sacks} also
developed a numerical code in which the real and imaginary part of the complex reflection coefficient are used as
input and the scattering optical potential is delivered as output. Fig.~\ref{fig5} shows the retrieved SLD of 
our sample which is determined by using the Sacks code.
One can see that the retrieved SLD corresponds to the mirror image of the unknown part of the sample with vacuum surroundings at both sides. 

\section{Conclusion}\label{conclusion}

Available neutron reflectometers are limited in the measurement direction of the polarization of the 
reflected beam according to which the polarization
of the reflected beam can only be measured in the same direction as the polarization of the incident beam. 
The polarization based theoretical approaches which deal
with the determination of the missing phase of reflection, suffer from this experimental deficiency. 
In this paper, we have optimized the method of reference layers
based on the one directional polarization analysis and resolved this problem. 

The optimized reference layer method deals with two different polarization measurements
in the same direction as the polarization of the incident beam. 
Moreover, the method resolves the the necessity of the change of experimental setup for polarization
measurements which is an important issue in experiments. 
In the optimized reference layer method, the information of the complex reflection coefficient is retrieved
only with polarization analysis. And in contrast to most of other method, there is no need to the measurement of reflectivity.
The method provides us with the modules
and phase of the complex reflection coefficient and results to the unique determination of the
scattering length density of the mirror image of the unknown sample.

In order to check the reliability of the method, a numerical simulation was performed and the knowledge of 
the missing phase of the reflection and as a consequence, the
SLD of the sample was truly retrieved. 
The numerical results certify that the method is reliable at this level and we look forward to its experimental implementations.


\begin{thebibliography}
\Latin
\bibitem{Xiao-Lin} Xiao-Lin Zhou, Sow-Hsin Chem, Phys. Rep. 257 (1995) 223.

\bibitem{Majkrzak1} C.F. Majkrzak, N.F. Berk, Physica B 336 (2003) 27.

\bibitem{Majkrzak2} C.F. Majkrzak, N.F. Berk, V. Silin, C.W. Meuse, Physica B 283 (2000) 248.

\bibitem{Majkrzak3} C.F. Majkrzak, N.F. Berk, Phys. Rev. B 58 (1998) 15416.

\bibitem{Majkrzak4} C.F. Majkrzak, N.F. Berk, Physica B 221 (1996) 520.

\bibitem{Majkrzak5} C.F. Majkrzak, N.F. Berk, U.A. Perez-Salas, Langmuir 19 (2003) 7796.

\bibitem{Majkrzak6} C.F. Majkrzak, N.F. Berk, Physica B 267–268 (1999) 168.

\bibitem{Masoudi1} S.F. Masoudi, A. Pazirandeh, Physica B 362 (2005) 153.

\bibitem{Masoudi2} S.F. Masoudi, M. Vaezzadeh, M. Golafrouz, G.R. Jafari, Appl. Phys. A 86 (2007) 95.

\bibitem{Leeb1} H. Leeb, J. Kasper, R. Lipperheide, Phys. Lett. A 239 (1998) 147.

\bibitem{Masoudi3} S.F. Masoudi, A. Pazirandeh, J. Phys.: Condens. Matter 17 (2005) 475.

\bibitem{Masoudi4} S.F. Masoudi, Eur. Phys. J. B 46 (2005) 33.

\bibitem{Leeb2} H. Leeb, E. Jericha, J. Kasper, M.T. Pigni, I. Raskinyte, Physica B 356 (2005) 41.

\bibitem{jahromi} S.F. Masoudi, S.S. Jahromi, Physica B 406 (2011) 2570–2573.

\bibitem{Gottfried} K. Gottfried, Quantum Mechanics (Benjamin, New York,1966), Chap. 40.

\bibitem{Chadan} K. Chadan, P.C. Sabatier, Inverse Problem in Quantum Scattering Theory, second ed., Springer, New York, 1989.

\bibitem{Sacks}  T. Aktosun, P. Sacks, Inverse Probl. 14 (1998) 211; T. Aktosun, P. Sacks, SIAM (Soc. Ind. Appl. Math.) J. Appl. Math. 60 (2000) 1340; T. Aktosun, P. Sacks, Inverse Probl. 16 (2000) 821.

\end{thebibliography}
\end{document}